# Self-Organization and The Origins of Life: The Managed-Metabolism Hypothesis

John E. Stewart

Evolution, Complexity and Cognition Group,
Center Leo Apostel, Vrije Universiteit Brussel,
Krijgskundestraat 33, B-1160 Brussels, Belgium.
Ph: +61422540984
future.evolution@gmail.com

**Abstract**: The 'managed-metabolism' hypothesis suggests that a 'cooperation barrier' must be overcome if self-producing chemical organizations are to undergo the transition from non-life to life. This dynamical barrier prevents un-managed, self-organizing, autocatalytic networks of molecular species from individuating into complex, cooperative organizations. The barrier arises because molecular species that could otherwise make significant cooperative contributions to the success of an organization will often not be supported within the organization, and because side reactions and other 'free-riding' processes will undermine cooperation. As a result, the barrier seriously limits the possibility space that can be explored by un-managed organizations, impeding individuation supported by complex functionality and the transition to life. The barrier can be overcome comprehensively by appropriate 'management'. Management implements a system of evolvable constraints that can overcome the cooperation barrier by ensuring that beneficial co-operators are supported within the organization and by suppressing free riders. In this way management can control and manipulate the chemical processes of a collectively autocatalytic organization, producing novel processes that serve the interests of the organization as a whole and that could not arise and persist spontaneously in an un-managed chemical organization. Management self-organizes because it is able to capture some of the benefits that are produced when its management of an autocatalytic organization enhances productivity by promoting cooperation. Selection will therefore favour the emergence of managers that take over and manage chemical organizations so as to overcome the cooperation barrier. The managed-metabolism hypothesis demonstrates that if management is to overcome the cooperation barrier comprehensively, its interventions must be digitally coded. In this way, the hypothesis accounts for the two-tiered structure of all living cells in which a digitally-coded genetic apparatus manages an analogically-informed metabolism.

**Graphical Abstract:**

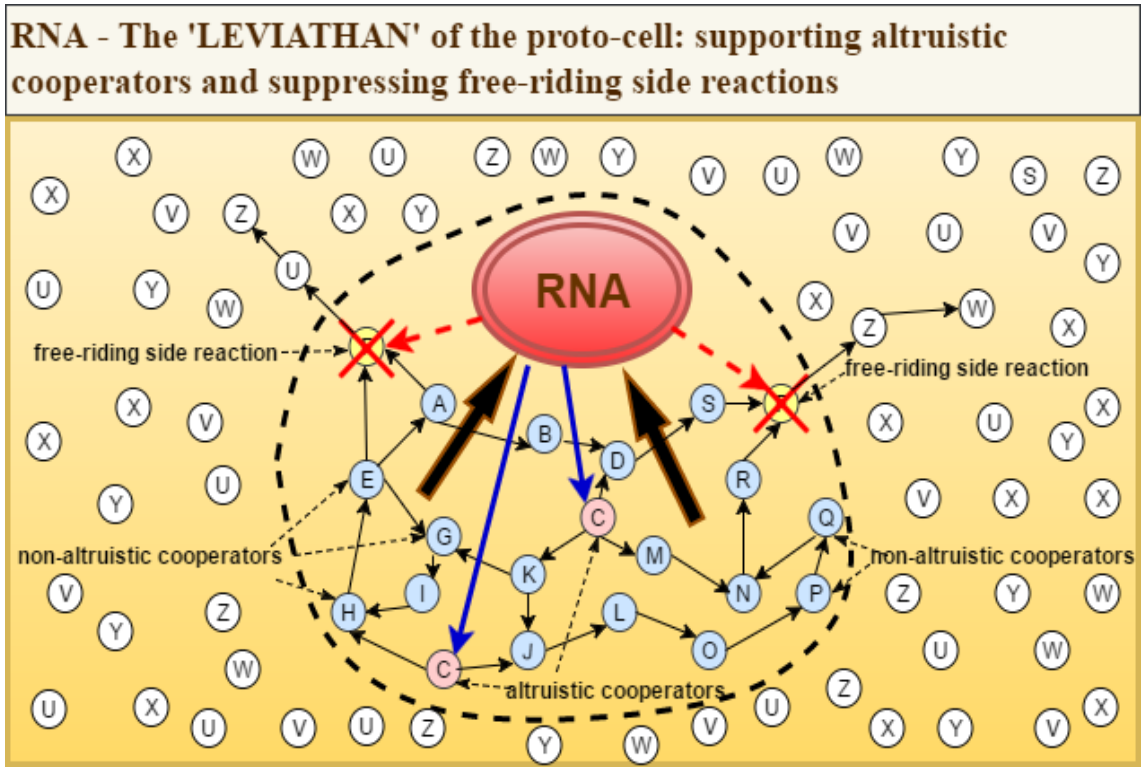

## 1. Introduction

The 'gene-first' hypothesis about the origins of life appears to continue to have significant support amongst many researchers [1-6]. Broadly, this hypothesis suggests that life began with the emergence of RNA molecules that had a capacity for both template-based self-replication and the ability to catalyse other reactions. As such, these molecules were capable of undergoing the standard evolutionary process: errors arising during copying of the molecules produced variant molecules; and variants that went on to reproduce more successfully (e.g. by catalysing reactions that enhanced their self-replication) tended to increase their representation in the population of molecules. According to this hypothesis, nothing more was needed: life as we know it was up and running.

On the surface, this hypothesis does not seem to demand anything more of the first genes and natural selection than what is achieved now by familiar gene-based evolutionary processes. It seems reasonable to assume that evolving RNA self-replicators would progressively discover and accumulate beneficial adaptations, just as organisms do now as they evolve. Under this hypothesis, the RNA replicators would progressively build around themselves increasingly complex metabolisms that would enhance their capacity to reproduce and survive. This model also seems to be consistent with the metaphor that self-replicating genes specify the 'blueprint' for the cell that contains them. If this metaphor accurately reflects the role of genes in a cell, it is easy to see how natural selection operating on variation in genes could gradually adapt and improve the blueprint to specify arrangements that are more adaptive.

However, on closer examination it is highly implausible that this proposed process could somehow clothe 'naked' self-replicating RNA molecules in a complex metabolism, starting with nothing. As we shall see, showing how this could happen is arguably a more difficult challenge than showing how such a self-replicating RNA molecule could arise from some organic soup in the first place (and this is proving to be a major challenge: so far, RNA replicators have not been shown to emerge even in organic soups that are designed, structured, manipulated and carefully constrained by teams of highly-qualified human chemists [e.g. see 7]).

The key implausibility of this aspect of the gene-first hypothesis is the proposition that naked RNA could progressively create around itself a highly complex, dynamical metabolism, starting from scratch and using an evolutionary mechanism that operates 'top down' through variation in RNA and generally makes only one small change at a time. No one has yet demonstrated in theory or practice that this is possible for a replicator that starts out 'naked' [8]. It is true that existing organisms including simple cells that contain genetic replicators are able to respond to selection with long sequences of adaptive change. But they begin with a supporting metabolism that potentially enables mutations to have significant effects. They don't start with nothing and build this complex enabling machinery from scratch. In this respect, the gene-first hypothesis seems even less plausible than an analogous 'government first' hypothesis of the origins of hierarchical human societies i.e. the proposition that these societies began with the emergence of 'naked' governments that then proceeded to somehow create around themselves all the rest of society (including economic and agricultural systems), starting from scratch. Such a hypothesis about human societies would be even more difficult to accept if governments were only able to establish and adapt governance through blind trial and error, as do genes.

Furthermore, a strong case can be mounted that that at least some of the metabolic and related constituents of cells have not been created by RNA in this way. There is evidence that the genetic apparatus of modern cells does not contain all the information embedded in the cell as a whole. As we shall see, the genetic apparatus does not include a blueprint for the building of a cell from scratch [9]. Even the simplest known cells contain information that is embodied in the cytoplasm and is not contained in the genetic apparatus [10]. As a consequence, the genetic apparatus could not reconstruct from scratch the processes and structures of the cell that embodies this information. This is highly suggestive that this information was not produced by RNA in the first place.

The evidence for this rests upon the fact that the processes and structures that are produced by a protein encoded by the genetic apparatus is not determined by the nature of the protein alone. Instead, it is determined by the interactions between the protein and the existing contents, processes and spatial structures of the cytoplasm. Cytoplasmic context is critically important in co-determining the effects of any protein coded by the genetic apparatus [10]. Information in the cell is therefore contained in both the genetic apparatus and cytoplasm.

This would not be such a problem for the 'gene-first' hypothesis if the genetic apparatus produces the cytoplasm from scratch. But it does not. Nor is it capable of doing so. In particular, there are many examples of super-molecular structures within cells that are never recreated from scratch within the cell. Instead they are replicated only by processes in which existing structures serve as scaffolding and templates for the production of new structures. In the absence of this scaffolding, the genetic apparatus cannot recreate the structures. Examples where this has been established include organelles within the cells such as the endoplasmic reticulum, cilia, mitochondria and the membranes that are part of many other structures [11-14]. It is clear that the genetic material in the cell is not at all like a blueprint for a building that specifies the location and structure of walls, windows, roof and floor and the spatial relations between them. It is now widely accepted in relation to eukaryote cells that some cellular organelles such as mitochondria had an evolutionary origin and history independent of the cells that now contain them [15]. It seems likely that other super-molecular structures now found in the simplest of know cells also had an evolutionary origin and history independent of the genetic apparatus.

What about other components of the metabolisms now embodied in the cytoplasm of cells, such as metabolic cycles and processes? Could they have also emerged and evolved to some extent independent of any self-replicating RNA? Whether they could is an issue of critical importance for understanding the origins of life. This is not just because it points to a viable alternative to the implausibility that metabolisms were progressively created from scratch by RNA. It is also because even if naked genes were somehow capable of building metabolisms, they would have been highly unlikely to do so if they could instead simply take over and manage metabolisms that had already emerged and evolved.

'Metabolism-first' hypotheses of the origins of life postulate that metabolisms in the form of organizations of molecular species indeed emerged first before self-replicating RNA. These organizations are envisaged as being self-producing because their constituent molecular species cooperate together to catalyse each other's formation and have access to sources of free energy and other necessary resources [16-23].

However, to what extent could the evolution of these metabolic organizations have eventually produced organizations that would qualify as living? Section 2 of this paper demonstrates that metabolic organizations would have encountered a 'cooperation barrier' that would prevent them from developing the complex individuality that I will argue is essential to the transition from non-life to life. This is analogous to the 'cooperation barrier' that is faced by cooperative organizations when they emerge at any level of organization of living processes. Section 3 examines the nature of the cooperation barrier that arises at these other levels, and Section 4 identifies how it has typically been overcome through the emergence of systems of evolvable constraints that are termed 'management' in this paper. Section 5 of the paper applies this understanding of management to the evolution of chemical organizations. It identifies how evolution is likely to have overcome the cooperation barrier facing proto-metabolisms by favouring the emergence of management in the form of RNA, thereby enabling the transition from non-life to life (the 'managed-metabolism hypothesis'). Section 6 examines the relationship between the managed-metabolism hypothesis and other hypotheses about the origin of life. It also considers how the managed-metabolism hypothesis could be developed further and tested.

It is worth emphasising at the outset that the managed-metabolism hypothesis is founded on a relational perspective about the nature of living processes: it considers that the dynamical relationships between the constituents of living processes are of paramount importance for understanding life (e.g. see [45]). In the main, the specific nature of the entities that constitute living organizations is significant only insofar as it influences the dynamical relationships that the entities can enter into, and the forms of organization in which they can participate. This relational perspective recognizes that there are a huge number of ways of organizing and combining the constituents of living processes that will not qualify as life. And there is only a comparatively infinitesimal number of ways of organizing them that will constitute life. A key goal of this paper is to identify the particular forms of organization that were able to transition from non-life to life.

## 2. The Cooperation Barrier and the Evolution of Metabolic Organizations

Reaction networks of molecular species that are self-producing because of their cooperative catalytic activity have been shown theoretically to exhibit some evolvability [21-23]. Simulations have shown that alternative autocatalytic reaction networks of organic polymers that coexist in the same environment can compete and undergo a form of natural selection. The heritable component of these reaction networks comprise viable cores which are self-sustaining collections of molecular species that catalyse each other's formation [21]. Competition and selection can occur between viable cores within a network as well as between networks that contain various combinations of different viable cores. Novel variation can arise between reaction networks in relation to viable cores through a number of processes, including by the acquisition of new cores through rare chemical events [21-22]. Compartmentalisation of reaction networks can enhance selection between networks and the accumulation of adaptations. Self-sustaining cooperative networks of organic polymers together with food sources and small-molecule autocatalytic cycles (e.g. pre-cursors of the reductive citric acid cycle with polymers catalysing steps in the autocatalytic cycle [22]) could conceivably evolve by these processes as proto-metabolisms.

However, the evolvability of these cooperative reaction networks is seriously limited by what I will refer to as a 'cooperation barrier'. As I will show in more detail in the next section, this dynamical barrier is analogous to the cooperation barrier that impedes the evolution of cooperative organization at all levels or organization, including amongst human and other multicellular organisms [24-26]. It turns out that the cooperation barrier facing self-producing organizations of molecular species must be overcome if the organizations are to make the transition from non-living chemical processes to life.

The nature of this barrier can be understood by considering an organization of molecular species that is self-producing because it is collectively autocatalytic—i.e. the formation of every species in the organization is catalysed by at least one other species, and the organization has access to appropriate sources of free energy and 'food' molecules. The organization contains a number of viable cores in which the formation of all molecular species is catalysed by all other members of the core. The cooperation barrier arises because molecular species that could contribute cooperatively to the survivability of a core and the organization as a whole may not be produced and sustained at an optimal level within the organization [24-26]. This can be the case irrespective of the significance of the contribution that these species could make to the success of the organization. How does such a barrier arise out of the dynamics of autocatalytic organizations? First, the formation of a beneficial molecular species might not happen to be catalysed by any other member of the organization (or it may not be catalysed at a level that is optimal for the organization). This is not likely to be uncommon—there is nothing at all in the nature of autocatalytic organization that guarantees that any particular molecular species that contributes positively to the organization will be catalysed in return. Second, it might occur where 'free-rider' molecular species and associated 'side reactions' take resources from the organization but do not contribute anything (or sufficient) in return (e.g. they do not catalyse the formation of other members of the organization). Free-riders can reduce the catalytic support, energy and material resources that might otherwise be available to members of the organization, undermining their ability to persist and contribute to the organization. Because free-riders do not use their resources to contribute to the organization, they may also out-compete those that do. The susceptibility of an organization to be undermined by free-riders is likely to increase as its complexity increases [27].

It is worth indicating here in more detail what is meant by the terms 'co-operator', 'unsupported co-operator' and 'free rider' at the level of chemical organisation, and relating these terms to the use of similar terms at the level of biological organisation. At the biological level, a co-operator organism is one which interacts with other organisms in ways which provide the others with fitness benefits e.g. by increasing the capacity of the others to reproduce successfully. Analogously, a co-operator molecular species is one which increases the rate at which other molecular species are produced in a chemical system by, for example, catalysing the formation of those molecular species. At the biological level, a supported (or non-altruistic) co-operator is one which can reproduce successfully in the population because it benefits sufficiently from its cooperative interactions with others to outweigh the fitness costs of its own cooperative actions. And an unsupported (or altruistic) co-operator is one which does not obtain sufficient benefits to reproduce successfully. Analogously, a molecular species which is a supported (non-altruistic) co-operator is one whose formation (reproduction) is supported sufficiently within the organisation (e.g. by catalysis from other molecular species) for it to be sustainable within the organisation. An unsupported (altruistic) molecular species is one which does not receive sufficient support within the organisation to be sustainable. At the biological level, a free-rider organism is one which receives fitness benefits from co-operator organisms, but does not provide fitness benefits in return. Analogously, a free-riding molecular species has its formation enhanced by co-operator molecular species (e.g. the co-operators might catalyse the formation of the free rider), but does not provide benefits to the co-operators in return (e.g. it does not catalyse the formation of other co-operators).

**Figure 1** is a schematic depiction of the architecture of a simple catalytic reaction network of molecular species which includes un-supported co-operators and free riders:

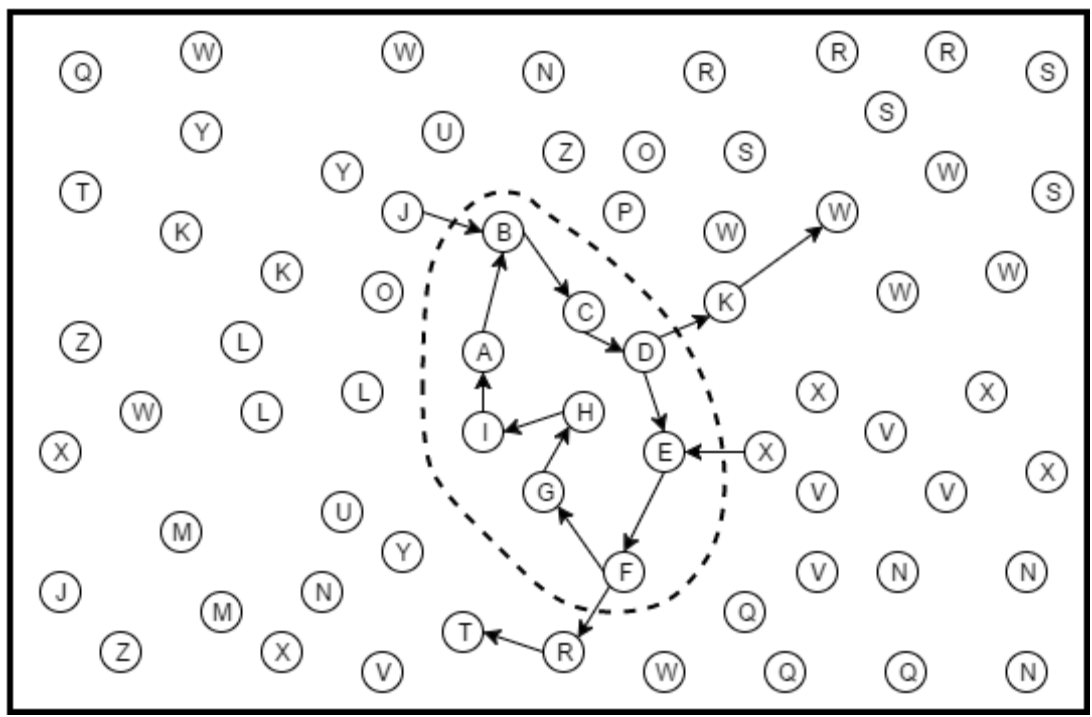

In **Figure 1**, each molecular species is represented by a circle containing a letter. The organizational architecture (excluding free-riders) is enclosed by a dotted line. The arrows between molecular species represent catalysis. The architecture shows that the formation of each member of the autocatalytic organization is catalysed by at least one other member. Molecular species J and X are unsupported (altruistic) co-operators: they contribute to the organization by catalysing the formation of members of the organization, but their own formation is not catalysed by any member of the organization. Molecular species K and R are free-riders: their formation is catalysed by members of the organization, but they don't contribute anything in return to the organization.

The cooperation barrier significantly limits the evolvability of autocatalytic organizations of chemical species. Organizations that are self-producing cannot include molecular species that could contribute significantly to the survivability of the organization but are not supported adequately within the organization. Such altruistically cooperative molecular species would not be produced within the organization and cannot persist as part of it (they are not dynamical attractors). Organizations that contain such species cannot be sustained or called into existence by selection, no matter how powerful the selection is, or how much the species would increase the competitiveness of the organization as a whole. Complex cooperative arrangements amongst molecular species that would produce highly advantageous supra-molecular structures, processes, sub-systems and systems could not evolve if any of the molecular species or processes that constitute them were unable to persist and be reproduced within the organization. Because of the unlikelihood that advantageous molecular species would coincidently be supported at an optimal level in the organization, molecular organizations that are subject to the cooperation barrier will be able to explore only a tiny proportion of the possibility space of logically conceivable organizations. As a consequence, the barrier seriously limits the extent of the possibility space that could be explored by catalytic reaction networks of molecular species, impeding their ability to evolve into complex, cooperative organizations. Organizations that were restricted by the cooperation barrier would not have been able to evolve the complex adaptive functionality that characterises all know living cells and constitutes their individuality (this paper takes the position that self-producing organizations do not qualify as living unless they exhibit individuality supported by complex functionality. For detailed discussion of the basis of this position, see 25, 28).

In order for an organization to have a comprehensive capacity to evolve beneficial cooperative functionality, any chemical species that would benefit the organization would have to be capable of being produced within the organization at a level that is optimal as circumstances change, and free riders would have to be able to be suppressed systematically. But chemical organizations that face the cooperation barrier fall far short of this capability. The overwhelming majority of molecular species that could contribute to the success of such organizations would not be produced at an optimal level. The chemistry that operates within the kinds of autocatalytic reaction networks that have been conceived to date cannot produce the complex cooperative organization that characterises life. For the transition to life, a new kind of chemistry was needed.

## 3. The Cooperation Barrier and Other Major Evolutionary Transitions

It is not only molecular organizations whose ability to explore the space of organizational possibilities is limited by a cooperation barrier. Cooperation barriers also impede the emergence of complex cooperative organization at each and every level of living organization [24-26]. It is therefore a barrier to the emergence of new levels of organization. For example, the cooperation barrier impeded the emergence of the cooperative organizations of eukaryote cells that became multicellular organisms, the organizations of organisms that became animal societies, the organizations of humans that became tribal societies, the organizations of human groups that became nation states, and is currently impeding the emergence of a complex, cooperative planetary entity. It should be noted that these emergences include many but not all of the major evolutionary transitions identified by Maynard Smith and Szathmáry [28] (e.g. it does not include sexual reproduction), and includes emergences that they do not (e.g. the emergence of a cooperative global organization [26]).

In order to work out how the cooperation barrier could have been overcome in the transition from non-life to life, it is useful to draw on the large body of research that has contributed to understanding why a cooperation barrier arises at other levels of organization and how it has been overcome at those levels.

A generalized agent-based approach can be used to understand how the cooperation barrier arises out of the dynamics of cooperative organizations at all levels [29]. Using this approach, agents represent the entities at each particular level (e.g. prokaryote cells, eukaryotes, multicellular organisms [including humans], tribes, nations etc.). Agents are capable of adaptation. Adaptation tends to maximize a function such as fitness or psychological utility. Agents may adapt by any process (e.g. including by processes as disparate as gene-based natural selection or psychological mechanisms). Agents are able to interact with each other in ways that may be adaptively advantageous. Cooperative organizations of agents will emerge where adaptations that constitute cooperative relationships between agents are also sufficiently beneficial to the individual agents themselves (e.g. where the adaptations provide net fitness or utility benefits to the individual agents that exhibit them). Where this condition is met, the relationships and the organization they constitute will persist and be reproduced through time. However, if agents fail to benefit sufficiently from cooperative interactions, the adaptations that underpin the cooperation will not be reproduced and persist, no matter how much the cooperation benefits the organization as a whole.

In many circumstances, individual agents will not benefit sufficiently from advantageous cooperative interactions, despite the potential of many forms of cooperation to significantly increase the net benefits available to the organization. This will be the case if co-operators fail to capture enough of the benefits they produce to outweigh the costs of their cooperation. As the huge body of research on cooperation referred to below has demonstrated, this failure can be expected to be commonplace. There is nothing in simple, unstructured forms of organization which guarantees that co-operator agents will always capture sufficient of the benefits they create. To the contrary, agents that support co-operators will tend to be outcompeted by agents that use resources only for their own benefit, without providing sufficient benefits to the organization in return (e.g. free-rider agents, including parasites, cheats and thieves). Free-riders will also tend to out-compete the co-operator agents themselves, and take resources that might otherwise support co-operators. Furthermore, there is nothing that guarantees that free-rider agents will always capture the 'harms' that they visit on the organization. For all these reasons, free-rider agents will tend to undermine complex cooperative organization.

As a consequence, the cooperation barrier will seriously restrict the possibility space of complex cooperative organization that can be explored at any level of organization. All forms of organization that include agents that provide significant net benefits to the organization but fail to capture sufficient of those benefits will not be able to persist (the sustained existence of their organizations will not be dynamical attractors).

## 4. Mechanisms that can Overcome the Cooperation Barrier

A huge literature attempts to identify particular mechanisms which enable co-operator agents to capture sufficient of the benefits they create to enable the emergence of some form of cooperative organization (e.g. see [26] for a brief overview). These mechanisms generally rely on co-operators capturing a disproportionate share of the benefits of cooperation because of the existence of circumstances which ensure they are disproportionately likely to interact with other co-operators. These biased patterns of interaction are typically produced by constraints that manifest as, for example: particular dispersal patterns; kin selection; group formation; compartmentalization; stochastic correction; other forms of population structure; pre-dispositions to cooperate preferentially with other co-operators; and pre-dispositions to punish and exclude free-riders. However, in general this body of research confirms the reality of the cooperation barrier. It has demonstrated that complex cooperative organization does not evolve readily. It has shown that simple cooperative relationships can emerge, but only in limited circumstances. Most researchers in this field would accept that the research has so far been unable to identify a general mechanism that could operate at all levels of organization and that would enable complex cooperative organization to emerge readily (e.g. see [28, 30]).

But the cooperation barrier has been overcome repeatedly and comprehensively during the evolution of life on this planet, enabling the emergence of complex cooperative organization at various levels. What mechanism(s) have enabled this? It is clear from the agent-based perspective sketched above that agents who provide significant net benefits to an organization would be able to persist if 'consequence-capture' applies—i.e. if agents capture sufficient of the benefits (and harms) they produce to sustain them at an optimal level in the organization. Comprehensive consequence-capture would massively expand the possibility space that can be explored by organizations at any level [29].

But what can enable consequence-capture? The emergence of what have been termed 'managers' can enable comprehensive consequence-capture within the organizations they manage [24-26, 29, 31]. Managers are powerful, evolvable agents (or coalitions of agents) that can control an organization to support co-operators and to suppress free riders. Managers control an organization by applying constraints. Constraints can influence the dynamical behaviour within the organization without being influenced in return (this is the essence of control). Constraints can operate to direct resources preferentially to co-operator agents, and can punish or suppress free-riders. In order to apply constraints, managers must function independently of the dynamical interactions within the organization proper. They must be able to stand outside and be able to act across the dynamic. The dynamical separation of a manager from an organization often results from the fact that the processes that constitute managers are larger in scale, involve slower rate processes and/or are relatively more stable than the processes that constitute the organization proper [32]. In order to survive and persist, managers do not depend on participation in exchange relations and other dynamical interactions within the organization (for a more detailed discussion of constraints and how they arise at any level of organization, see [32]). Instead, managers can use their constraining power to appropriate whatever resources they need from the organization. Without the capacity to constrain (to influence without being influenced in return), any attempt by managers to appropriate resources for themselves or to distribute resources to particular agents could be undermined by other agents, and free riders could escape control. Just as powerless members of a human organization are unable to control or manage the organization, powerless agents within an organization cannot apply constraints to it or begin to manage it—they cannot influence without being influenced in return. Because Ryan et al [46] fail to take into account this relationship between power and constraints, they fail in their attempt to provide a comprehensive explanation of how the major evolutionary transitions in individuality were enabled. In particular, they fail to justify the feasibility of their key assumption: they assume that a single allele could somehow empower otherwise ordinary members of a population to apply pro-social constraints that suppress free-riders (including by punishing them), and then maintain cooperation in the face of thieving and other forms of free-riding. Furthermore, they fail to explain why selection would favour individuals who use power pro-socially, rather than individuals who use their power to increase their fitness in more direct ways e.g. by simply appropriating benefits produced by others.

**Figure 2** is a schematic representation of the architecture of an externally-managed organization of agents:

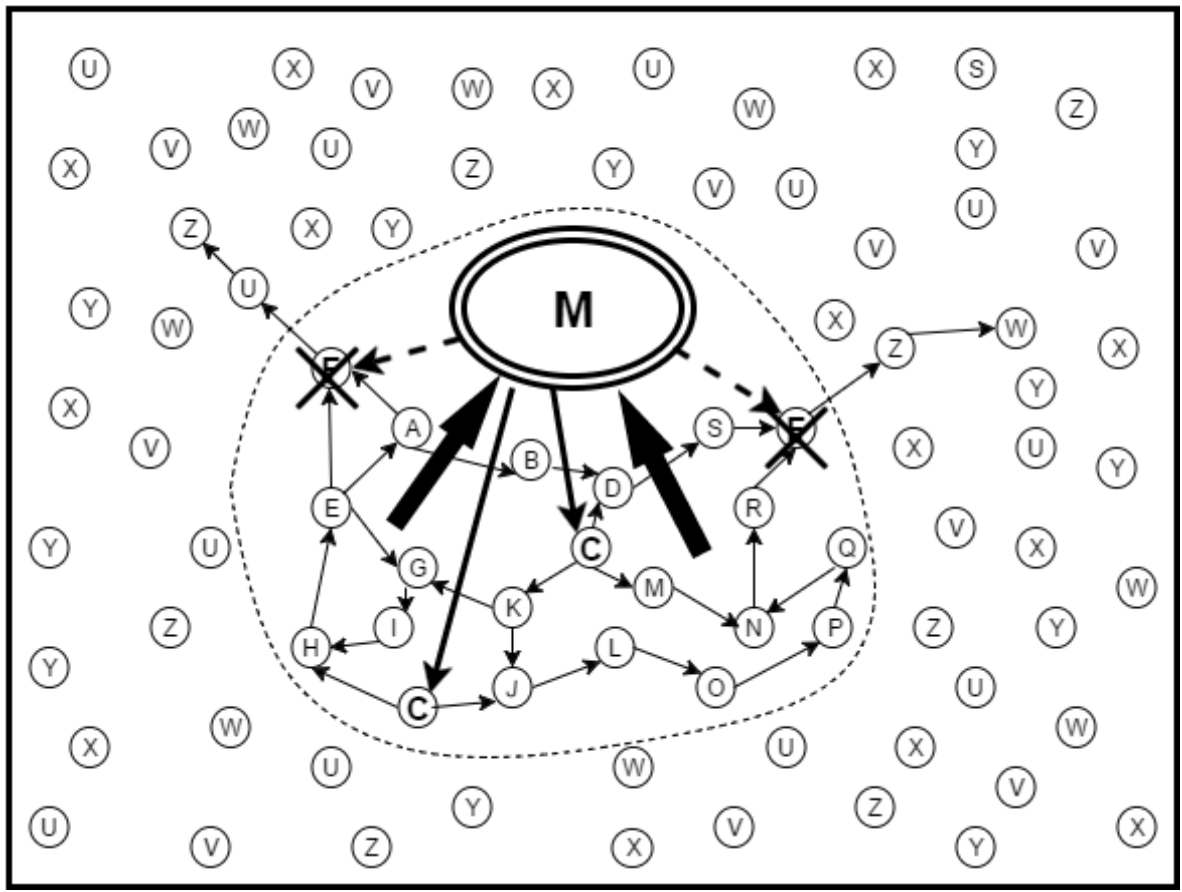

In **Figure 2**, each agent is represented by a circle containing a letter. The organizational architecture is enclosed by a dotted line. M represents the powerful, evolvable manager. The part of the organization that does not include the manager is comparable to the autocatalytic architecture depicted in figure 1. The two agents marked '**F**' are free-riders on that organization, and the two agents marked '**C**' are co-operators that contribute to the organization but are not supported in return. The normal arrows represent the flow of benefits within the organization. The bolded arrows originating from the manager represent support by the manager for the two co-operators (**C**) that are not otherwise supported within the organization. The two dashed and bolded arrows originating from the manager represent the suppression by the manager of free-riders (**F**). Finally, the heavily bolded arrows that point towards the manager represent the unilateral appropriation by the manager of benefits from the organization.

Management and the constraints it applies can be more or less enabling or more or less prescriptive. Furthermore, where management itself is comprised of a coalition of agents, it will encounter its own cooperation barrier. This barrier can be overcome by constraints that suppress competition within management. Management can be external to the agents that are being managed, or can be internal to the agents and distributed across them [24, 31, 25]. Examples of 'external management' include: RNA/DNA management of the metabolism of a cell; management of a human society by its government; and management of the employees of a corporation by the board of directors (in these three examples, the managers may seem to be integrated parts of the organization. However, on closer examination the power relationships between the manager and other members of the organization are unmistakeable). Examples of 'distributed internal management' include: a multi-cellular organism in which the behaviour of cells is controlled across the organization by genetic constraints that are reproduced in each cell; and a human tribal society in which the behaviour of each member is constrained by internalized norms as well as genetic constraints that are reproduced in each member. In the case of internal distributed management, the behaviour of every agent in the organization is controlled and coordinated by a system of constraints that is reproduced within each and every agent. As such, the constraints reach across the entire organization, and also capture the benefits (and harms) produced by their impacts on the organization as a whole. Distributed internal management can be as effective at controlling an organization as external control. But where it operates, it is often mistaken for an absence of control [34].

Stewart [31, 25-26] examines in some detail how the coincidence of interests between management and the organization as a whole drives the self-organization and emergence of management (management can appropriate greater resources from an organization that is managed in ways that overcome the cooperation barrier).

From the broader perspective developed here, the huge literature on the emergence of cooperation can be seen as a search for particular circumstances in which constraints that allow some degree of consequence-capture just happen to exist. But 'nature' has not limited itself to producing advantageous cooperative organization only in those special circumstances where suitable constraints exist as 'happy accidents'. Instead, in the transition to life and in all subsequent major transitions, 'nature' has incorporated within organizations a mechanism that has the capacity to search for and implement whatever systems of constraints will enable consequence-capture and the emergence of complex cooperative organizations. Evolvable management enables the discovery and implementation of whatever sets of constraints will maximize appropriate consequence-capture in any organization in any situation.

## 5. The Managed-Metabolism Hypothesis and the Origins of Life

### 5.1 The Emergence of Management

As is the case at other levels of organization, appropriate management had the potential to overcome the cooperation barrier which was faced by self-producing organizations of molecular species. But what would have driven the emergence of management in the transition from non-life to life? And what characteristics would management need to have in order to manage molecular organizations effectively?

The prime candidates for early managers are evolvable coalitions of polymers such as RNA that had the power to intervene catalytically in organizations of molecular species [24, 31, 25-26]. Initially, RNA or other coalitions might simply have plundered the contents of collectively autocatalytic organizations, using them to assist their own reproduction and then moving on to plunder other organizations (e.g. by using as 'food' contents of an autocatalytic organization that are constituents of the RNA molecules). Importantly, these coalitions would not have participated in the internal catalytic interactions and relationships that occurred within the organizations they exploited. The coalitions would have stood outside the organizations dynamically and appropriated the resources they needed (for more detail on the nature of this relationship, see [32]. The capacity to do this, together with their evolvability and their potential to catalytically intervene in organizations unilaterally, would have given them the *potential* to control and manage an organization as a proto-metabolism.

What would cause these RNA or other coalitions to realize this potential to become managers? Why would they use their power to overcome the cooperation barrier? What would drive the transition from plunderers to managers? Coalitions could achieve an advantage if they discovered ways to use their evolvable catalytic capacities to enhance the productivity of an organization and manage it as a proto-metabolism. This is because coalitions that could increase the productivity of organizations could be able to harvest greater benefits from them for their own use. The existence of the cooperation barrier provided an enormous potential for coalitions to increase their fitness by doing just this [24, 25-26, 47]. The coalitions could discover ways to intervene in organizations to support molecular species that contribute to the productivity of the organization, but would not be supported otherwise. Furthermore, they could suppress side-reactions and other free-riders that impede productivity (e.g. by degrading the first catalyst in a chain of side reactions, or by preferentially supporting alternative processes within the organization that do not produce side reactions).

Selection would favour coalitions of RNA that managed chemical organizations in ways that increased the benefits they could harvest from them. As a result, coalitions would increasingly move away from plundering and destroying organizations. Coalitions would more and more become effective managers, with each coalition managing a particular organization as a proto-metabolism, enhancing the productivity of the organization and increasing the resources that the coalition could harvest on an on-going basis. A coincidence of interests would arise between the coalition and the proto-metabolism it managed. This evolutionary sequence is broadly analogous to the historical transition which was undergone by Mongol tribes: they began as plunderers that destroyed other societies and then moved on to new conquests and pillaging. But eventually the Mongols became rulers of the societies they conquered, introducing systems of governance (management) that enhanced the productivity of the societies. Rather than plunder a society once, they could harvest an enhanced stream of benefits from it on an on-going basis.

The development of management capabilities gave RNA 'the power of life and death' over proteins that were supported or inhibited by its catalytic capacities. RNA had the power to determine whether or not these proteins could exist as members of the organization. The seminal book on evolutionary transitions by Maynard Smith and Szathmáry [28] did not recognise the critical importance of the use of power by managers to enable the transition from non-life to life and the transitions at most other levels of organization. Maynard Smith and Szathmáry instead argued that the use of power in this way is largely restricted to the human level of organization in the form of armed force and/or a consensus imposed by a majority. Although it is true that RNA was not armed, it could develop considerable power over an organization and it was this power that enabled it to overcome the cooperation barrier.

## 5.2 The Transition from Chemistry to Life.

Effective, evolvable management (whether RNA or otherwise) would have enabled self-producing organizations to transition from non-living chemistry to life. As we have seen, un-managed, self-producing chemical organizations were only able to explore a possibility space that is seriously limited. But effective, evolvable management would have changed everything. It opened up enormous new areas of possibility space to self-producing organizations, enabling them to go far beyond what is possible through un-managed chemical interactions and processes. Management opened the door to entirely novel and hitherto unknown arrangements of matter that were self-producing. It did so by controlling and manipulating chemical processes so that they served the organization's functions and purposes. The nature and functioning of the constituents of the organization were no longer determined by chemistry alone. It was now dictated by the evolutionary needs of the organization as a whole. With comprehensive consequence-capture, the constituents of self-producing organizations would tend to adapt in ways that served the interests of the organization. As a consequence, managed organizations would tend to evolve and adapt as coherent wholes that could develop all the characteristics of individuality. In contrast, un-managed autocatalytic organizations were like ecosystems—they could contain autocatalytic cycles and processes but would not evolve as individuals (comprehensive management and consequence-capture are prerequisites for the full emergence of individuality). In the service of their individuality, managed organizations would explore an extensive new space of possible organizational forms, relationships, processes and subsystems. These could not arise through normal chemical processes in the absence of management. With the transition to life, a new kind of chemistry emerged on the planet: managed chemistry. From a thermodynamical perspective, management enabled the emergence of entirely new kinds of material processes that could dissipate energy gradients more effectively. Management was the key to the transition from non-life to life.

From this perspective, the central function of the DNA apparatus (and RNA before it) was not the storage of information. Its primary significance in the evolution of life was to serve as management that enabled the cooperation barrier that separates chemistry from life to be overcome. The storage of information is incidental to the primary function of the DNA/RNA apparatus which is to manage. Effective management requires memory.

### 5.3 What kind of management would be favoured by selection?

As we have seen, individual selection operating on managers would tend to have favoured management that overcame the cooperation barrier facing proto-metabolisms. But what were the particular characteristics that management would need to enable it to overcome the barrier fully as possible and be favoured by selection? What were the forms of management that would have had the potential to manage most effectively?

#### 5.3.1   The limitations of managers comprised only of autocatalytic networks

The primary function of management is to overcome the cooperation barrier by supporting molecular species and processes that are beneficial to the organization and by suppressing side reactions and other free riding processes that are not. Ideally, effective managers would need to be able to catalyse the formation of any metabolic polymer that could benefit the organization. Managers that were limited in their capacity to catalyse useful reactions in a metabolism would tend to be less effective than those that were not so limited. Would this requirement be able to be met by managers that were themselves comprised only of autocatalytic networks of polymers in which the polymers did not self-replicate individually through a template-based process (i.e. by networks of polymers that are replicated/reproduced only collectively)? It is conceivable that suitable autocatalytic networks of such 'managerial polymers' might have some capacity to manage a proto-metabolism that included networks of other polymers (metabolic polymers) and small-molecule autocatalytic loops and processes. This might be the case where some of the members of the managerial autocatalytic network were able to catalyse at least some beneficial reactions in the proto-metabolism. However, such autocatalytic networks would be very limited in their management capabilities. This is because their evolvability would be severely restricted for two reasons: first, they would not be able to evolve and explore possibility space through a 'copying-with-errors' process. Second, like all un-managed autocatalytic networks, they would encounter a cooperation barrier in full. This cooperation barrier would limit them to exploring only a small proportion of the space of catalytic possibilities and therefore of the space of beneficial management interventions. There would be many conceivable managerial polymers that could catalyse beneficial reactions in the proto-metabolism that would not be sustainable in the autocatalytic network due to the cooperation barrier. Because such managerial networks would not be able to catalyse the formation of many potentially useful metabolic polymers, they would be incapable of implementing many management controls that would be beneficial to any proto-metabolism they might manage.

For example, consider a manager constituted by an autocatalytic network (coalition) of RNA molecules that do not self-replicate individually by a template-based process. The RNA manages a protein-catalysed proto-metabolism. Because of the cooperation barrier faced by this managerial network, there would be many RNA molecular species that would not be sustainable in the network—their formation would not be catalysed by other members of the network. This would be likely to include many RNA molecules that could catalyse the formation of particular proteins that would be useful in the proto-metabolism, but that are not sustainable within the proto-metabolism because of the cooperation barrier it faces. These metabolic proteins might, for example, catalyse beneficial reactions or help constitute useful structures, if they were able to persist in the proto-metabolism due to support by appropriate managerial RNA. For these reasons, managers constituted only by autocatalytic networks of RNA or other polymers are unable to overcome comprehensively the cooperation barrier faced by proto-metabolisms.

### 5.3.2   The superiority of digitally-coded management

Given the seriously restricted management capabilities of managers comprised only of autocatalytic networks of polymers that are incapable of self-replication through a template-based process, is there a different kind of molecular system that does not share these limitations? We will see that if a molecular system is to manage proto-metabolisms comprehensively, it must have at least two characteristics:

(a) In principle, it must have the potential to generate a relatively unlimited range of interventions in the organization it manages.

(b) The success of any given variant intervention must depend only on its contribution to the organization as a whole and not, for example, on its ability to compete internally with other variants.

As we have seen, the first of these requirements cannot be met by managers comprised only of autocatalytic networks of managerial polymers that are replicated only collectively and are incapable of template-based self-replication. However, it could be satisfied by a managerial polymer that self-replicates individually through a copying process that produces occasional errors. In principle, the replication and mutation of such a managerial polymer would be able to explore fully the space of all possible combinations of monomers that could be included in that kind of managerial polymer (this assumes that the managerial polymer is associated with catalytic arrangements that enable all variant polymers to also self-replicate individually. However, although the relaxation of this assumption changes the details of the argument advanced below, it does not change its conclusions about the advantages of digitally-coded management). Because such a managerial polymer could (in principle) give rise to any possible polymer of its kind, it would have the potential to discover and exploit all possible catalytic effects that could be produced *by that kind of polymer*. For example, such a manager comprising RNA polymers would (in principle) be able to discover and utilize any RNA polymer that had a beneficial catalytic effect on some aspect of the proto-metabolism it manages e.g. by catalysing the formation of a useful protein enzyme. Of course, in practice, many polymers would not be discovered in any particular population of managed proto-metabolisms. This is because, for example, relevant mutations might never arise, or because a particular managerial polymer may only be able to be reached through a sequence of mutations that are not each viable. Again, however, this does not disturb the central thrust of the argument being advanced in this section.

However, would the ability to produce any kind of managerial polymer enable comprehensive management, at least in principle? Consider a metabolism that includes an autocatalytic network of metabolic polymers such as proteins. Would a manager comprising self-replicating RNA polymers be able to comprehensively overcome the cooperation barrier for the network of those metabolic polymers? To do so, ideally the manager would have to be able to catalyse at an optimal level the formation of any possible metabolic polymer. But it would be unable to do this. Although the manager could (in principle) produce any possible molecule of the *managerial* polymers that constitute it, there might be useful *metabolic* polymers whose formation was not catalysed by any of these.

How might a manager be constituted so that it could catalyse the production of any possible metabolic polymer, in principle? The manager could do so if managerial polymers served as templates for the production of metabolic polymers such that the sequence of monomers was determined by the sequence of monomers in the managerial polymers. Since the mode of reproduction of managerial polymers (copying-with-errors) could, in principle, produce any variant of managerial polymer with any sequence and length of monomers, such a translation process could (in principle) produce any possible metabolic polymer. Consider again the example of a manager comprised of RNA polymers that self-replicate and serve as templates for the production of protein metabolic polymers. In this arrangement, the sequence of monomers in the RNA polymer would act as a digital code for the sequence of monomers in the proteins [34, 10]. In principle, it could provide the basis for a system that is able to produce any feasible protein at an optimal level through time.

The second requirement for comprehensive management is that the success of any variant intervention must depend only on its contribution to the organization. This condition would be violated if the managerial polymers could self-replicate independently and therefore compete with each other. If this occurred, a variant that produced a metabolic polymer that was highly beneficial to the organization might be out-competed and unable to persist within the manager or unable to produce the metabolic polymer at an optimal level. This is a *much more limited* version of the cooperation barrier that is faced by autocatalytic networks, and it could be overcome far more simply. In particular, competition between managerial polymers could be prevented if all the polymers were constrained so that they were only able to replicate together, as a unity. For example, all managerial polymers could be bound together to form a single entity (e.g. a single chromosome) that replicates as a unit [35] (this is analogous at the level of human organization to a single king who rules a society and is bound by strict succession arrangements [25]). As a further example which encompasses standard mitosis and meiosis, managerial polymers might be bound together to form a small number of entities (again, for example, chromosomes) that are then constrained by additional arrangements which ensure that they replicate together only when the cell reproduces, and then only as a unit [36] (this is analogous at the level of human societies to the constitution and associated arrangements that constrain a parliament [25]).

In summary, managers constituted only by autocatalytic networks of non-self-replicating polymers would be unable to overcome the cooperation barrier faced by proto-metabolisms that they manage. This is because the cooperation barrier encountered by the managerial network itself would prevent it supporting many processes that would be highly beneficial to the proto-metabolism. However, this limitation would not apply to a manager constituted by self-replicating polymers which were themselves able to catalyse metabolic polymers such as proteins. In principle, such a manager would be able to explore the space of all possible managerial self-replicators. But there is no guarantee that the managerial self-replicators would have the potential to catalyse all possible metabolic polymers (e.g. RNA falls far short of having the potential to catalyse the formation of all possible proteins). Only a manager constituted by self-replicating polymers whose sequence of monomers serves to specify the sequence of monomers in metabolic polymers could achieve this (in principle). Hence managers must utilize a digital code if they are to overcome comprehensively the cooperation barrier faced by 'analogical' proto-metabolisms of molecular species. 'Analogical' managers are unable to do so. In addition, further arrangements are necessary to constrain managerial self-replicators to prevent competition between them on the basis of criteria other than the success of the interventions they initiate.

For these reasons, the transition from non-life to life had to await the emergence of management that embodied such a digital code. Before digitally-coded management emerged, layers of less effective and less evolvable forms of 'analogically-informed' management was likely to have emerged, forming hierarchies of management (this may have included layers of management constituted by autocatalytic networks of non-self-replicating RNA as well as networks of peptides/proteins or other polymers). The eventual takeover of analogically-informed metabolisms by digitally-informed management is analogous at the human level to the procedural redescription that occurs during human development in which procedural knowledge is re-described and extended by declarative knowledge [37].

The digitally-informed arrangements embodied in the genetic apparatus were subsequently co-opted by evolution to overcome the cooperation barrier at other levels of organization e.g. to provide the distributed internal management that underpinned the emergence of multi-cellular organizations and that also underpinned the emergence of organizations of multi-cellular organisms such as insect societies. It was not until the emergence of complex human societies that a new kind of digital code evolved in the form of language [10].

## 6. Relationship Between the Managed-Metabolism Hypothesis and Other Hypotheses About the Origins of Life

### 6.1 Other Hypotheses

The managed-metabolism hypothesis incorporates a number of the key elements of other major hypotheses about the origins of life and combines these with new features which overcome the deficiencies of these other hypotheses.

In particular, like gene-first hypotheses, the managed-metabolism hypothesis relies upon the emergence of self-replicating RNA molecules. But unlike gene-first hypotheses, it does so far more plausibly. Many versions of the gene-first hypothesis postulate that RNA arose spontaneously from an unstructured organic-rich soup. In contrast, the managed-metabolism hypothesis proposes that the emergence of RNA molecules (or other potential managers) was scaffolded by the prior emergence of an 'ecosystem' of autocatalytic networks and cycles of molecular species (for a discussion about the likely emergence of a community of proto-organizations, see [38]). As these networks and cycles self-organized and evolved, they are likely to have built up chemical processes and species that significantly increased the likelihood that RNA could emerge. Dyson [17] outlines one specific process by which this might have occurred. Of course, even though the first RNA molecules might have emerged from self-producing chemical organizations of other molecular species, this does not mean that RNA would have always thereafter been dependent upon those organizations for its survival and reproduction. Rather, as discussed above, its evolvable catalytic capacity gave it the potential to unilaterally appropriate resources from other organizations and eventually to develop the capacity to manage them.

In another major difference from most gene-first hypotheses, the managed-metabolism hypothesis does not propose that 'naked' self-replicating RNA molecules proceeded to progressively create around themselves a complex, supporting metabolism, starting from scratch. Instead, it argues that potential managers are much more likely to have taken over and managed pre-existing organizations that emerged in the chemical 'ecosystem', rather than to have created them afresh (particularly given the difficulties of building highly complex, dynamical organizations from scratch using an evolutionary mechanism that operates 'top down' and generally makes only one small change at a time).

Like metabolism-first hypotheses about the origin of life, the managed-metabolism hypothesis also relies on the emergence of autocatalytic networks of molecular species. But current metabolism-first hypotheses do not include any mechanism that would overcome the cooperation barrier sufficiently to enable the emergence and persistence of highly complex metabolisms (it has been shown that compartmentalization and selection operating at the level of compartments can enable the emergence of some cooperation, particularly by the suppression of free-riders [e.g. see 39, 28]. But it has not been demonstrated that this mechanism can account for the emergence within self-producing molecular organizations of the complex cooperative organization that characterizes life). The managed-metabolism hypothesis proposes that the emergence of the complex, cooperative metabolisms found in modern cells required the emergence of comprehensive management that was able to support co-operators at optimal levels and to suppress free-riders.

A widely considered hypothesis that has some surface similarities to the managed-metabolism hypothesis is Dyson's parasite/symbiosis hypothesis of the origins of life [17]. Dyson suggests that RNA emerged first within self-producing chemical organizations and developed a parasitic relationship with them. He goes on to argue that this relationship eventually co-evolved into a mutually-beneficial symbiotic relationship (he suggests that this parallels the later symbiotic origin of the eukaryote cell identified by Margulis [15]). However, Dyson's hypothesis misses both the fundamental reason why evolution favours the role played by RNA in the transition to life as well as the very nature of that role. More specifically, he does not recognise the existence of a cooperation barrier which (a) seriously limits the ability of self-producing proto-metabolisms to develop advantageous cooperative arrangements and (b) creates the potential for RNA to provide significant advantages to itself and to the organization it manages by overcoming the barrier (these are absences that are shared by all other hypotheses that postulate an RNA or DNA takeover of metabolisms).

Dyson did raise the possibility that autocatalytic metabolisms may encounter some of the 'catastrophes' that have been shown by simulations to beset RNA quasi-species and hypercycles [40]. These catastrophes include elements of the cooperation barrier that I have described. However, Dyson left to future simulations an assessment of whether any of these catastrophes would indeed apply to proto-metabolisms. His hypothesis therefore did not come close to recognising the role that RNA had in overcoming these and other instances of the cooperation barrier, or how the significant benefits that flow from this could drive the comprehensive take-over of proto-metabolisms by RNA.

Because Dyson's hypothesis misses the role of RNA in overcoming the cooperation barrier, it also misses: (a) the critically important power relationship between the RNA and the proto-metabolism that enables the RNA to emerge as an evolvable manager; (b) that the power of RNA management to apply constraints across the proto-metabolism enables it to progressively overcome the cooperation barrier (the RNA apparatus becomes 'the Leviathan' of the proto-cell [for more about how 'the Leviathan' overcomes the cooperation barrier in human societies, see [41]); and (c) that the integration of simple cells into emerging eukaryote cells is, in fact, an example of the capacity of powerful management to overcome the cooperation barrier (as are all other relevant major evolutionary transitions). It is not an example of mutually-beneficial symbiosis between equals (the genetic apparatus of the emerging eukaryote cell manages/enslaves the simple cells that are incorporated into it [e.g. see 28, 25]). So it is the managed-metabolism hypothesis that is consistent with the other relevant major evolutionary transitions, not the parasite/symbiosis hypothesis.

Ganti's 'Chemoton Model' of a simple cell also has some superficial similarities to elements of the managed-metabolism hypothesis. Like the managed metabolism hypothesis, the Chemoton Model includes a 'genetic subsystem' that controls and regulates the dynamics of the system as a whole (e.g. see [42]). However, like Dyson, Ganti does not recognize anything like the cooperation barrier which, according to the managed-metabolism hypothesis, is the key driver of the emergence and evolution of management. Also unlike the managed-metabolism hypothesis, Ganti's approach does not suggest that the Chemoton's genetic subsystem emerges to support co-operators and suppress side reactions and other free-riders. In their extensive discussion of the Chemoton Model, Griesemer and Szathmary acknowledge that it does not provide a solution to the side-reaction problem [43].

**6.2 Testing the Hypothesis**

Because the managed-metabolism hypothesis includes elements that are also part of other hypotheses about the origins of life, theoretical and empirical work on those elements of other hypotheses will also assist in testing and developing parts of the managed-metabolism hypothesis. For example, simulations and laboratory studies that demonstrate the plausibility of the self-organization of self-producing organizations of molecular species will be highly relevant to the plausibility of comparable aspects of the managed-metabolism hypothesis, as will demonstrations that their evolvability is impeded by a cooperation barrier. Any laboratory demonstration that adds to the plausibility of the view that RNA could emerge 'spontaneously' in particular circumstances will also significantly strengthen the managed-metabolism hypothesis. But unlike many 'gene-first' scenarios, the plausibility of the managed-metabolism hypothesis would be even more significantly enhanced by any demonstration that the emergence of RNA is far more likely if it could be scaffolded by the kinds of chemical processes that are likely to have arisen only in self-producing proto-metabolisms.

However, any testing of the hypothesised emergence of management and its potential to overcome the cooperation barrier is likely to require theoretical and experimental work that is specifically focused on the managed-metabolism hypothesis. It seems likely that this work would need to begin with attempts to simulate the emergence of management and its take-over of proto-metabolisms, rather than through laboratory work. Until un-managed proto-metabolisms can be produced in the laboratory, attempts to produce actual managed proto-metabolisms is likely to be premature. Theoretical and laboratory work on the emergence of management at the level of chemical organizations can also be informed and enhanced by work on the emergence of management at other levels of organization, including at the level of human organizations (e.g. see [25], [44]).

## 7. Conclusions

Un-managed organizations of autocatalytic networks of molecular species can be self-producing and can also evolve by natural selection to a limited extent. But un-managed organizations encounter a cooperation barrier. The barrier limits their capacity to develop complex cooperative arrangements within the organization. This in turn seriously limits the emergence within organizations of complex functionality that serves the interests of the organization as a whole and enables it to function and adapt as a coherent entity. Complex individuality therefore cannot emerge. All known living cells exhibit individuality supported by complex functionality. If we take this to be an essential criterion for life, un-managed self-producing organizations are unable to make the transition from non-life to life.

For complex individuality to evolve, digitally-coded management is necessary to overcome the cooperation barrier. If the cooperation barrier did not exist and if un-managed self-producing autocatalytic organizations of molecular species were somehow able to evolve complex functionality, living processes would be organized entirely differently to the way they are: organizations of molecular species would be able to individuate fully and transition from non-life to life without the emergence of management. If this were the case, living cells would not have their distinctive two-tiered structure comprising an analogically-informed metabolism governed by a digitally-coded management.

However, the cooperation barrier does exist. Management was essential for the transition from non-life to life. Effective management unleashed the cooperative creativity that produced 'endless forms most beautiful and most wonderful'. Without management, natural selection would have been incapable of producing or shaping life of any form.

The relational perspective that underpins the managed-metabolism hypothesis recognises that there may be many chemical species and processes other than those found in life on earth that can constitute life. This is because these alternatives are capable of engaging in the cooperative relationships and forms of organization that qualify as life—i.e. they can constitute organizations that are self-producing, digitally-managed and capable of developing individuality supported by complex functionality (provided, of course, they have access to suitable sources of free energy and other resources).

On this planet, proteins, RNA and DNA play key roles in living processes. Metabolisms in cellular forms of life are constituted by particular chemical cycles and processes that are catalysed primarily by proteins, and digitally-coded management is constituted by particular forms of RNA and DNA. At higher levels of organization on this planet, self-producing organizations and digitally-coded management are constituted by larger-scale entities (e.g. by humans that use language and by organizations of humans at the level of societies). On other planets and in other circumstances, the actual constituents of the simplest living processes may be quite different to those on Earth. But they are likely to be organized into the two-tiered structure we find here: a cooperative metabolism organized by digitally-coded management.

## 8. Acknowledgements

The author acknowledges the benefit of useful discussions with David Richards, Mark Roddam and Wilson Kennel.